\documentclass[11pt]{iopart}

\usepackage{graphicx}
\usepackage{epsf}
\usepackage{longtable}
\usepackage{hyperref}
\usepackage{graphics,epsfig,placeins,subfigure,wrapfig}
\expandafter\let\csname equation*\endcsname\relax

\expandafter\let\csname endequation*\endcsname\relax
\usepackage{amsmath,amssymb,calc,amsfonts}
\usepackage[usenames]{color}
\usepackage{soul}
\usepackage{amsfonts}
\usepackage[mathscr]{euscript}

\usepackage[toc]{appendix}

\usepackage{harmony}

\DeclareMathAlphabet\mathbfcal{OMS}{cmsy}{b}{n}

\def\eadnew#1#2{\address{#2 E-mail: \mailto{#1}}} 

\begin{document}

\title{Newman-Janis Ansatz for rotating wormholes}

\author{ A. C.  Guti\'errez-Pi\~neres  $^{1,2,  \checkmark}$,  N. Hern\'andez $^{1,*}$  and  C. S. Lopez-Monsalvo $^{3,**}$ }

\address{$1$  Escuela de F\'isica, Universidad Industrial de Santander, A. A. 678, Bucaramanga 680002, Colombia.}
\address{$2$  Instituto de Ciencias Nucleares, Universidad Nacional Aut\'onoma de M\'exico,  \\AP 70543,  M\'exico, DF 04510, M\'exico.}
\address{$3$  Conacyt-Universidad Aut\'onoma Metropolitana Azcapotzalco Avenida San Pablo Xalpa 180,  Azcapotzalco, Reynosa Tamaulipas, 02200 Ciudad de M\'exico, M\'exico.}

\eadnew{acgutier@uis.edu.co}{$^{\checkmark}$}
\eadnew{nicolas.hernandez27@correo.uis.edu.co}{$^{*}$}
\eadnew{cslopezmo@conacyt.mx}{$^{**}$}


\begin{abstract}
A central problem in General Relativity is obtaining a solution to describe the source's interior counterpart for Kerr black hole. Besides, determining a method to match the interior and 
exterior solutions through a surface free of predefined coordinates remains an open problem.
In this work, we present the ansatz formulated by Newman-Janis to generate solutions to the Einstein field equation inspired by the mentioned problems.
We present a collection of independent classes of exact interior solutions of the Einstein equation describing rotating fluids with anisotropic pressures. Furthermore, we will elaborate 
on some obtained solutions by alluding to rotating wormholes.  
\end{abstract}
\section{Introduction}
In 1963 \cite{kerr1963gravitational} Roy Kerr discovered their solution to the Einstein field equations describing a spinning black hole. One year later, Newman and Janis  \cite{newman1965note}  demonstrated another form to derive this solution by making a complex transformation to the Schwarzschild solution. Since then, there is an increasing interest in generating new solutions from a prescribed seed using the 
 original Newman and Janis ideas.  

Nowadays, the discussion about  the Newman-Janis Ansatz (NJA) mathematical nature remains open to research.  As far as we know,  maybe the Drake  and  Szekeres work \cite{drake2000uniqueness} and the set by Azreg \cite{azreg2014static}  are currently the most telling about the mathematical view and the applicability of the NJA. In 2016, one of us and 
H. Quevedo \cite{gutierrez2016newman}  asked themselves if conformal symmetry prevails after using this formalism. The answer was no for a conformastatic seed. This paper shall continue this investigation and explore the possibility of finding new solutions to the Einstein equations describing self-rotating fluids. 

Since the papers by Ellis in 1973 \cite{ellis1973ether} and Morris and  Thorne in 1988  \cite{morris1988wormholes} set up an aesthetical and conceptual renaissance of the notion of wormholes, abundant literature has grown up proposing new examples of non-traversable and traversable wormholes. Three among the many problems concerning the original idea remain open to debate: First, how to match the interior solution of the Einstein equations, whose metric is given by a wormhole, to an exterior solution. Second, to provide wormholes formalism with arbitrary symmetries. Moreover, finally, to construct rotating wormholes with appropriate physical behavior.  For a discussion on matching wormholes to their exterior space-time, see \cite{lemos2003morris}, whereas for discussion on rotating wormholes, see \cite{teo1998rotating, azreg2016wormholeA, azreg2016wormholeB}.
 
In section \ref{sec: NJA} of this paper,  we present the ansatz formulated by Newman-Janis to generate solutions to the Einstein field equations for rotating space-times. 
We employ the differential geometry in the fashion that lets transparent the idea according to which the formalism abandons the transformation between metric spaces in favor of mappings between tetrads.

Section \ref{sec:RAFNJA} presupposes that the rotating solution obtained using  NJA  describes a stationary interior space-time corresponding to an anisotropic fluid without heat. We implement this assumption in the Einstein field equations and achieve five independent classes of interior solutions.

Finally, in section \ref{sec:NJARW}, we present a short description of the mathematical aspects necessary to generate the rotating counterpart of static wormholes by employing the NJA. Next, we construct the Kerr-Newman wormholes using the Reissner-Nordstr\"om wormhole as a seed solution to illustrate the discussed ideas.  We conclude this work with some remarks and open proposals for further investigations.
 \section{The Newman-Janis Ansatz} \label{sec: NJA}
 Let $(t, r, \theta, \phi)$ be the coordinates corresponding to point $p$ in a manifold $\cal M$, and  $( {\partial_{t} },\,{\partial_{r}},\, {\partial_{\theta}},\,{\partial_{\phi}})$ the basis tangent 
vector  of  $T_p({\cal M})$. Thus,  we have a  general metric tensor  $ \mathbf{{\cal G}_0} \in T^*_p({\cal M})$ given in the  manner by
                         \begin{align}\label{eq:GeneralStaticMetric}
                                            \mathbf{ {\cal G}_0} &= G(r) \operatorname{d}{t} \otimes  \operatorname{d}{t} 
                                            - \frac{1}{F(r)}  \operatorname{d}{r} \otimes  \operatorname{d}{r}
                                            -  H(r) ( \operatorname{d}{\theta} \otimes  \operatorname{d}{\theta}   
                                           + \sin^2 {\theta} \operatorname{d}{\phi} \otimes  \operatorname{d}{\phi} ) \ .
                             \end{align} 
 Below, we outline the procedure employed for the first time in 1964  by Newman and Janis  \cite{newman1965note}  to derivate the Kerr metric  by performing a  complex coordinate   
 transformation on the Schwarzschild solution.  We sketch it in a modern fashion consisting of foursteps and employ a static space-time sufficiently general to generate a master metric describing an axially symmetric rotating space-time.                     
                         
 \subsection{First step}
 We  perform the  mapping from  spherical coordinates to  outgoing Eddington-Finkelstein coordinates. In what follows, we shall use this metric tensor as  a seed solution to implement the Newman Janis Ansatz  (NJA).   To this end, we first  perform the transformation 
                              $$ \mbox{\Acht} : (t, r, \theta, \phi )  \longrightarrow (u, r, \theta, \phi )\ , $$
which maps spherical coordinates to  outgoing Eddington-Finkelstein coordinates through the relation
                                   $$ u = t - \int{ \frac{dr}{\sqrt{F G}}} \ .$$                                                                                                
Consistently, the mapping  $\mbox{\Acht}$ induces the pullback
                                    $ \mbox{\Acht}^{*}  $ 
which  carries  metric    $\mathbf{{\cal G}_S}$  into  metric   $ \mathbf{{\cal G}_0}$, i. e.,  $ \mbox{\Acht}^{*}  \mathbf{{\cal G}_S}  =  \mathbf{{\cal G}_0}$ .  
Then, the metric (\ref{eq:GeneralStaticMetric}) can be get through the action of the  map $ \mbox{\Acht}^{*}$ on   the metric $  \mathbf{{\cal G}_S}  \in T^*_{\mbox{ \Acht}(p) } ({\cal M})$
  given by
                              \begin{align}\label{eq:Eddington-Finkelstein_StaticMetric}
                                                 \mathbf{{\cal G}_S} &= \mathbf{E^t} \otimes  \mathbf{E^t}  - \mathbf{E^r} \otimes  \mathbf{E^r} 
                                                - \mathbf{ E^{\theta} } \otimes  \mathbf{  E^{\theta} } 
                                                -   \mathbf{E^{\phi}} \otimes  \mathbf{E^{\phi}}\ ,
                               \end{align}      
where the  basis vectors  $\mathbf{E^i}$  are given by the orthonormal tetrad
                                   \begin{align*}
                                                       \mathbf{E^t}  & \equiv  \sqrt{G} \operatorname{d}{u}  + \frac{ \operatorname{d}{r} }{\sqrt{F}},\\
                                                       \mathbf{E^r}   & \equiv   \frac{\operatorname{d}{r}}{ \sqrt{F}} , \\
                                                       \mathbf{ E^{\theta} }   & \equiv   \sqrt{H}  \operatorname{d}{\theta} , \\ 
                                                        \mathbf{E^{\phi}} & \equiv  \sin {\theta}  \sqrt{H}   \operatorname{d}{\phi} . 
                                    \end{align*}                  
 \subsection{Second step}
We can turn any real vector space ${V}$ into a complex vector space ${V}^{\mathbb{C}}$ by forming  the set  ${V} \times {V} $  of all pairs 
                                $(\mathbf{ E_{i}}, \mathbf{ E_{j}})$, with $\mathbf{ E_{i}}\ , \mathbf{ E_{j}} \in {V}$  
and then writing 
                               $(\mathbf{ E_{i}}, \mathbf{ E_{j}})$
 as 
                               $\mathbf{ E_{i}} + i\, \mathbf{ E_{j}}$. 
With this definition  ${V}^{\mathbb{C}}$  becomes a complex vector space (For more details, see \cite{de1992relativity} ). Hence, we can construct  a set of vectors forming a basis of 
                                      $\{  \mathbf{L} ,  \mathbf{N} ,  \mathbf{M} , \mathbf{W}  \}$ in $T_p^{\mathbb{C}}({\cal M})$
defined by
                                    \begin{align*}
                                                        \mathbf{L}   &  \equiv  \frac{1}{\sqrt{F}}\, (     \mathbf{E_t}  +   \mathbf{E_r}  ) 
                                                                            =   \, \partial_r  , \\
                                                         \mathbf{N}  &  \equiv  \frac{\sqrt{F}}{2} \, (     \mathbf{E_t}   -    \mathbf{E_r}  )  
                                                                            =   \sqrt{\frac{F}{G}}\,  \partial_u  - \frac{F}{2}\, \partial_r    , \\
                                                        \mathbf{M}   &  \equiv   \frac{\sqrt{2}}{2} \, (     \mathbf{E_{\theta}}  + i    \mathbf{E_{\phi}}  ) 
                                                                            =     \frac{\sqrt{2}}{2 \sqrt{H}} \, \partial_{\theta}  +    \frac{\sqrt{2}\,i}{2 \sin {\theta} \sqrt{H}} \, 
                                                                                     \partial_{\phi}  , \\
                                                        \mathbf{W}  &  \equiv   \frac{\sqrt{2}}{2}\,  (     \mathbf{E_{\theta}}   - i   \mathbf{E_{\phi}}  ) 
                                                                            =     \frac{\sqrt{2}}{2 \sqrt{H}} \, \partial_{\theta}  -    \frac{\sqrt{2}\,i}{2 \sin {\theta} \sqrt{H}} \,
                                                                            \partial_{\phi} \, .
                                                          \end{align*} 
This basis constitutes a complex null tetrad, i.e., consists of two real null vectors $\mathbf{L}$, $\mathbf{N}$ and two complex conjugate null vectors 
$\mathbf{M}$ , $\mathbf{W}$  thus, the scalar products of tetrad vectors satisfy:
                                $  \mathbf{L} \cdot  \mathbf{L}  =    \mathbf{M} \cdot  \mathbf{ M}  =      \mathbf{N} \cdot  \mathbf{ N} 
                                 =      \mathbf{L} \cdot  \mathbf{ M}  =    \mathbf{M} \cdot  \mathbf{ N} =0$    and  $  \mathbf{L} \cdot  \mathbf{N}  
                                 = -   \mathbf{M} \cdot  \mathbf{ W}  =1 . $  
 \subsection{Third step (a) }
Next, we introduce the ``rotated'' Eddington-Finkelstein coordinates by performing the mapping 
                           $${\mbox{\Sech}_R} : (u, r, \theta, \phi)  \longrightarrow  (u_R, r_R, \theta_R, \varphi_R) $$
 through the transformation 
                                                   \begin{align}
                                                           u_R & = u - i a \cos{\theta}, \nonumber\\
                                                           r_R  &=  r +  i a \cos{\theta}, \nonumber\\
                                                          \theta_{R} & = {\theta}, \nonumber\\
                                                            \phi_R & =  \phi.
                                                             \end{align} 
The map ${\mbox{\Sech}_R}$ naturally  induces the push-forward $(\mbox{\Sech}_R)_{*}$,
                  $$ ( \mbox{\Sech}_R)_{*}  : T_p^{\mathbb{C}}({\cal M})   \longrightarrow  T_{\mbox{\Sech}_R(p)}^{\mathbb{C}}({\cal N}) \ .$$
Hence, it is very easy to  verify that the basis vector 
                                   $ \mathbf{\xi} \equiv  \{    \mathbf{L},    \mathbf{N},    \mathbf{M},    \mathbf{W}  \}  \in T_p^{\mathbb{C}}({\cal M}) $
is mapped  by $\mbox{\Sech}_R$ into the basis 
                        $ \mathbf{\xi}_R \equiv \{ \mathbf{L}_R,    \mathbf{N}_R,    \mathbf{M}_R,   
                                 \mathbf{W}_R  \} \in T_{{\mbox{\Sech}_R}(p)}^{\mathbb{C}}({\cal N})$
and then                                                    
                             \begin{align}\label{eq: EFR_tetrad}
                              \mathbf{L}_R & =  \partial_{r_{_R}}, \nonumber\\
                                \mathbf{ N}_R & = \sqrt{\frac{B}{A}} \partial_{u_{_R}} - \frac{B}{2} \partial_{r_{_R}}, \nonumber\\
                                  \mathbf{ M}_R & = \frac{\sqrt{2} }{2\sqrt{\Psi}} (i a \sin{\theta_{_{R}}} \partial_{u_{_R}}  
                                  - i a \sin{\theta_{_{R}}}  \partial_{r_{_R}} 
                                    + \partial_{\theta_{_R}}   +  \frac{i}{ \sin{\theta_{_{R}}}}  \partial_{\phi_{_R}} ), \nonumber\\
                                    \mathbf{W}_R & = \frac{\sqrt{2} }{2\sqrt{\Psi}} (-i a \sin{\theta_{_{R}}} \partial_{u_{_R}}  
                                    + i a \sin{\theta_{_{R}}}  \partial_{r_{_R}} 
                                   + \partial_{\theta_{_R}}   -  \frac{i}{ \sin{\theta_{_{R}}}}  \partial_{\phi_{_R}} ),              
                                    \end{align} 
 where $A= A(r, \theta), \,B=  B(r, \theta),\, \Psi= \Psi(r, \theta)$   are functions to be known.
  \subsection{Third step (b)}
As it is evident from (\ref{eq: EFR_tetrad}), $\mathbf{\xi}_R$ is a complex null consisting of  two real null vectors $\mathbf{L}_R, \mathbf{N}_R $ and  two complex conjugate 
null vectors $\mathbf{ M}_R, \mathbf{ W}_R $, hence, the scalar products of tetrad vectors satisfy:
                                $  \mathbf{L}_R \cdot  \mathbf{L} _R =    \mathbf{M}_R \cdot  \mathbf{ M}_R  =      \mathbf{N}_R \cdot  \mathbf{ N}_R
                                 =      \mathbf{L}_R \cdot  \mathbf{ M}_R  =    \mathbf{M}_R \cdot  \mathbf{ N}_R =0$    and  $  \mathbf{L}_R \cdot  \mathbf{N}_R 
                                 = -   \mathbf{M}_R \cdot  \mathbf{ W}_R  =1 . $  
Moreover, 
                                   \[       \eta^{ij}_R = 
                                                          \left(
                                                                   \begin{array}{cccc}
                                                                      0 & 1 & 0 & 0 \\
                                                                      1 & 0 & 0 & 0 \\
                                                                      0 & 0 & 0 & -1 \\
                                                                      0 & 0 & -1 & 0 \\
                                                                      \end{array}
                                                                                                     \right)
                                         \]
  are  the components of a  specific metric tensor  $ \mathbf{ {\cal G}}_{_{R } } $ respect to the complex null tetrad, i.e., 
                             \begin{align} 
                             (\mathbf{ {\cal G} }_{_{R } }  )_{i j}  =    (\mathbf{L}_{_{R } } )_{i }  (\mathbf{N}_{_{R } }  )_{j } +  (\mathbf{N}_{_{R } }  )_{i }  (\mathbf{L}_{_{R } }  )_{j } 
                                             -    (\mathbf{M}_{_{R } } )_{i }  (\mathbf{W}_{_{R } }  )_{j } -  (\mathbf{W}_{_{R } }  )_{i }  (\mathbf{M}_{_{R } }  )_{j } .
                               \end{align}
Hence, the metric tensor $\mathbf{g}_{_{R }} $ can be expressed in terms of the real basis $\partial_R $ as:
                             \begin{align}\label{eq:PreGeneralStationaryMetric}
                             \mathbf{ {\cal G} }_{_{R }  }   &  = A{\operatorname d}{u_{_R}} \otimes {\operatorname d}{u_{_R}} 
                                        +  2 \sqrt{\frac{A}{B}} {\operatorname d}{u_{_R}} \otimes {\operatorname d}{r_{_R}} 
                                        +  2 a \sin^2 \theta_R \left( \sqrt{\frac{A}{B}} - A\right) {\operatorname d}{u_{_R}} \otimes {\operatorname d}{\phi_{_R}} \nonumber\\
                                     &  - 2 a \sin^2 \theta_R  \sqrt{\frac{A}{B}}  {\operatorname d}{r_{_R}} \otimes {\operatorname d}{\phi_{_R}} 
                                           - \Psi  \ {\operatorname d}{\theta_{_R}} \otimes {\operatorname d}{\theta_{_R}}   \nonumber\\    
                                     &     +  \frac{\sin^2 \theta_R }{B}   \left[  a^2\sin^2 \theta_R  \left( AB - 2 \sqrt{AB} \right)  
                                     - B\Psi \right]{\operatorname d}{\phi_{_R}} \otimes {\operatorname d}{\phi_{_R}}           .                                                                    
                                  \end{align} 
  \subsection{Fourth step}
We introduce the Boyer-Linquist  coordinates by the  map  
                                          $$ \mbox{\Zwdr}_{BL} :  (u_R, r_R, \theta_R, \phi_R) \longrightarrow  (T, R, \Theta, \Phi)  $$
            \begin{align*}
                u_R  = T - \frac{a^2{\sqrt{G} + {\sqrt{F}H}}}{(a^2 + F H) \sqrt{G} } R \ , \qquad
                 \phi_R  =  \Phi  + \frac{a}{a^2 + F H}  R \ .
                   \end{align*}        
Thus, the metric (\ref{eq:PreGeneralStationaryMetric}) admits the form                                                           
                  \begin{align}\label{eq:GeneralStationaryMetric}
                    \mathbf{ {\cal G} }   &  = \frac{  (a^2\cos^2\theta + F H) G \Psi}{(a^2\cos^2\theta \sqrt{G} + \sqrt{F} H )^2} {\operatorname d}{T} \otimes {\operatorname d}{T}                  
                          - \frac{\Psi}{FH + a^2} {\operatorname d}{R} \otimes {\operatorname d}{R}
                         \nonumber\\
                          &  - \Psi {\operatorname d}{\Theta} \otimes {\operatorname d}{\Theta} - \frac{2 (F \sqrt{G} -  \sqrt{F} ) a \sin^2{\Theta} \Psi H \sqrt{G}}{(a^2\cos^2\theta \sqrt{G} 
                          +  \sqrt{F} H  )^2} {\operatorname d}{T} \otimes {\operatorname d}{\Phi}  \nonumber\\
                      & - \Psi\sin^2\Theta \left[ 1 + \frac{a^2\sin^2\Theta H\left ( 2  \sqrt{\frac{F}{G}}     +  \frac{a^2\cos^2\Theta}{H}  -  F  \right) }{ (a^2\cos^2\Theta \sqrt{G} +  \sqrt{F} H )^2}  \right] 
                         {\operatorname d}{\Phi} \otimes {\operatorname d}{\Phi} \ .
                             \end{align}                    
This result shows that, by implementing the former successive step, the Newman-Janis Ansatz generates a stationary axially symmetric space-time from a static spherically symmetric.                                                              
  \section{Rotating anisotropic fluids from the  NJA}\label{sec:RAFNJA}
To interpret the tensor metric (\ref{eq:GeneralStationaryMetric}),  we presuppose that it describes a stationary interior space-time corresponding to an 
anisotropic fluid without heat. This kind of fluids admits a representation in terms of a tensor's components in the following form,
                  \begin{align}\label{eq: EMT-anisotropic_fluid}
                                     (\mathbf{T})^{ab} = \mu \ (\mathbf{V})^{a} (\mathbf{V})^{b} + P_R\  (\mathbf{e}_R)^{a} \ (\mathbf{e}_R)^{b} 
                                                                 + P_{\Theta}\  (\mathbf{e}_{\Theta})^{a} \ (\mathbf{e}_{\Theta})^{b} 
                                                                 + P_{\Phi}\  (\mathbf{e}_{\Phi})^{a} \ (\mathbf{e}_{\Phi})^{b} ,
                                                                 \end{align}
provided that the space time endowed with some Lorentzian  pseudo-Riemannian metric with signature $(+\, 2)$  
                                                                          \footnote{
                                                                         We developed most of the following calculations using a computational routine in Maple 2021. 
                                                                         To preserve the consistency with the Differential Geometry package, we changed the signature 
                                                                         $(-\,2)$ to $(\,2)$  in this section (and the rest of the work).
                                                                         }
                     \begin{align}
                                        (\mathbf{{\cal G} })^{ab} =  - \mathbf{V}^{a} \mathbf{V}^{b} 
                                                                                   + \  (\mathbf{e}_R)^{a} \ (\mathbf{e}_R)^{b} 
                                                                                   +  (\mathbf{e}_{\Theta})^{a} \ (\mathbf{e}_{\Theta})^{b} 
                                                                                  +   (\mathbf{e}_{\Phi})^{a} \ (\mathbf{e}_{\Phi})^{b} ,
                                                                                  \end{align}
where $\mu(R, \Theta)$ is the energy density of the fluid and $P_{R}(R, \Theta)$, $P_{\Theta}(R, \Theta)$ and $P_{\Phi}(R, \Theta)$ are the pressure along the 
corresponding directions. The word-lines of the fluids are  integral curves of the velocity vector  $\mathbf{V}$, which      satisfies the condition
 $ \mathbf{g} ( \mathbf{V},  \mathbf{V}   ) = -1$. Here,  $(\mathbf{V})^a$,  $(\mathbf{e}_{R})^{a}$,  $(\mathbf{e}_{\Theta})^{a}$ and  $(\mathbf{e}_{\Phi})^{a}$ are  
 the components of the vectors of the orto-normal tetrad given by
                            \begin{align}
                              \mathbf{V}  & =  - \frac{1}{\sqrt{\Psi (F H + a^2)}}\left( \left(\sqrt{ F}H/\sqrt{G} + a^2 \right){\partial}_{T}  + a\, {\partial }_{\Phi}  \right) \ ,\nonumber\\
                                             \mathbf{e}_R & = { \sqrt{\Psi(F H + a^2)} }/{ {\Psi} }\ {\partial}_{R} \ ,\nonumber\\
                                     \mathbf{e}_{\Theta} & =  \frac{1}{ \sqrt{\Psi}}\ {\partial}_{\Theta} \ ,\nonumber\\
                                        \mathbf{e}_{\Phi}  & =  - \frac{1}{\sqrt{\Psi} \sin{\Theta} }\big( a \sin^2(\Theta){\partial}_{T}  +  {\partial}_{\Phi} \big) \  . \end{align}
Thus,  the Einstein field equations $\mathbf{G}= 8 \pi  \mathbf{T}$ with the metric tensor  (\ref{eq:GeneralStationaryMetric}) and the energy-momentum 
tensor given by  (\ref{eq: EMT-anisotropic_fluid}) yields the following system of independent equations for the energy density and the anisotropic pressures:
                                  \begin{align}\label{eq:dynamic-variable1}
                                                      \mu\  = \ &  \frac{1}{32\pi\rho^4\Psi^3}\Big\{ 3\rho^4(\Psi^2_{,\Theta}+\Delta\Psi^2_{,R}) - 4\rho^4\Psi(\Psi_{,\Theta\Theta}
                                                                    +  \Delta\Psi_{,RR})-  \rho^2(\Psi^2)_{,R}\big(\rho^2\Delta_{,R}\nonumber\\
                                                                    &-a^2\sin^2\Theta\; K_{,R}\big)+2a^2\sin^2\Theta \rho^2 \Psi^2 K_{,RR}-2\cot\Theta\;\rho^2(\Psi^2)_{,\Theta}(K+a^2)\nonumber\\
                                                                 & +  \Psi^2\big[3 a^2\sin^2\Theta K^2_{,R}\;-4(\rho^2+4a^2\cos^2\Theta)(K+a^2)+12\cos^2\Theta\;\Delta^2\big]\Big\} \ ,        
                                                                \end{align} 
                        \begin{align}\label{eq:dynamic-variable2}
                                               P_{R} \ =  \  &  \frac{1}{32\pi\rho^4\Psi^3}\Big\{4\rho^4\Psi\Psi_{,\Theta\Theta}+3\rho^4(\Psi^2_{,R} - \Psi^2_{,\Theta})
                                                                  +  \rho^2(\Psi^2)_{,R}\big(\rho^2\Delta_{,R}-2\Delta K_{,R}\big)\nonumber\\
                                                              &  +  \rho^2(\Psi^2)_{,\Theta}\big[a^2\sin2\Theta+2\cot\Theta(K+a^2)\big]-4K(K+2a^2)\nonumber\\
                                                              &  +  \Psi^2\big[a^2\sin^2\Theta\:K^2_{,R}+4a^2\cos^2\Theta\left(6(\rho^2+a^2\sin^2\Theta)-(\Delta+a^2\cos^2\Theta)\right)\big]\Big\} \ ,
                                                            \end{align}
                       \begin{align}\label{eq:dynamic-variable3}
                                            P_{\Theta} \ = \ &\frac{1}{32\pi\rho^4\Psi^3}\Big\{4\rho^2\Delta\Psi(\rho^2\Psi_{,RR}-\Psi K_{,RR})+2\rho^4\Psi^2(F_{,RR}H+FH_{,RR})\nonumber\\
                                                             & +  2\rho^2\big[\cot\Theta\,(\Psi^2)_{,\Theta}\big(K+a^2(1+\sin^2\Theta)\big)+(\Psi^2)_{,R}(\rho^2\Delta_{,R}-\Delta K_{,R})\big]\nonumber\\
                                                             & +  3\rho^4(\Psi^2_{,\Theta}-\Delta\Psi^2_{,R})+\Psi^2\big[K^2_{,R}(8\Delta-a^2\sin^2\Theta)-6\rho^2K_{,R}\Delta_{,R}\nonumber\\
                                                             & +  4(\rho^4F_{,R}H_{,R}+a^2\cos^2\Theta\;\Delta)\big]\Big\} \ ,
                                                            \end{align}
                                       \begin{align}\label{eq:dynamic-variable4}
                                           P_{\Phi} \ = \ &\frac{1}{32\pi\rho^4\Psi^3}\Big\{4\rho^4\Psi(\Psi_{,\Theta\Theta}+\Delta\Psi_{,RR})+2\rho^4\Psi^2(F_{,RR}H+FH_{,RR})\nonumber\\
                                           &-3\rho^4(\Psi^2_{,\Theta}+\Delta\Psi^2_{,R})
                                                         + 2\rho^2\Delta\big(\cot\Theta\,(\Psi^2)_{,\Theta}-\Psi^2K_{,RR}\big)+\Psi^2\big[3K^2_{,R}(2\Delta\nonumber\\
                                                         &-a^2\sin^2\Theta)-6\rho^2\Delta_{,R}K_{,R}+ 4\rho^4F_{,R}H_{,R}+4\Delta(2a^2\cos^2\Theta-K)\big]\Big\} \ .
                                                              \end{align}
where $`` , "$ denotes derivative and we have used the simple notation  $\rho^2\equiv K + a^2\cos^2(\Theta)$, $K \equiv \sqrt{F}H/ \sqrt{G}$ and   $\Delta \equiv FH + a^2$. 
In addition to the above system,  the Einstein  equations give us the following two non-trivial independent equations: 
                                  \begin{align} \label{eq: constrains1}
                                                     \big( 3\Psi_{,R}\Psi_{,\Theta}   -  2\Psi_{,R\Theta}\Psi \big)\rho^4 & + 3 a^2\sin(2\Theta) K_{,R}\,\Psi^2\,  = 0 \ , 
                                                       \end{align}
                                 \begin{align} \label{eq: constrains2}
                                                       \big[ (\Psi K_{,R})_{,R}   + 2\Psi_{,\Theta}\cot(\Theta) \big] ( K + a^2\cos^2(\Theta))^2 & - \Psi\big(2K-a^2\cos^2(\Theta) + K^2_{,R}\big) = 0 \ ,
                                                           \end{align}
  which we must solve to determine the previous dynamic quantities fully characterizing the fluid. Hence, a large number of solutions corresponding 
to rotating anisotropic fluids arose after getting solutions to the system of equations (\ref{eq: constrains1}) - (\ref{eq: constrains2}). As we know, there are at least five classes of solutions of the above equation system. They are:
    \begin{align*}
         & \text{Class\ 1}:     \  \   K  = c_1, \qquad \Psi = \frac{a^2 \cos^2{\Theta + c_1}}{\cos{\Theta}} \ , \\
         &      \text{Class\ 2}:     \  \   K   = R^2 \ , \qquad \Psi = \frac{8c_1^2(a^2 \cos^2{\Theta + R^2)}}{ (\sqrt{2} c_3\cos{\Theta} R - c_2)^2}   \ , \\
           &          \text{Class\ 3}:     \  \   K  =(R-c_1)^2,\qquad \Psi =-\frac{c_2^2\big[a^2\cos^2\Theta + \big(R-c_1\big)^2\big]}{\big(R-c_1\big)^2\cos^2\Theta} \ ,  \\
            &     \text{Class\ 4}:     \  \   K  = R^2 + c_1R+c_2\ , \qquad \Psi = a^2\cos^2\Theta + R^2 + c_1R + c_2\ \ , \\
             &        \text{Class\ 5}:     \  \   K  =-\frac{\big(e^{c_2+R}+2\big)^2}{2\,e^{c_2+R}},\qquad 
                                                            \Psi = \frac{2\,a^2\cos^2\Theta \; e^{c_2+R}-\big(e^{c_2+R}+2\big)^2}{\cos^2\Theta\big(e^{c_2+R}+2\big)^2} \ ,
     \end{align*}
$  c_1 ,  c_2  \ $  and  $\  c_3 $  are arbitrary constants, and $ K^2  = H^2{{F}/{G}} $ is a relation between the coefficients of the seed static metric.  As we can see, each class defines  a kind of specific rotating fluid. In particular, if we introduce $c_1 = c_2$ in  Class 4,   we get  a solution of the form $\Psi= a^2\cos(\Theta)^2 + R^2$, and $K=R^2$ form. Additionally, if  we choose $K \equiv \sqrt{F/G}H$ determining the Schwarzschild solution 
 (i.e., $H = R^2$ and $F = G = 1- \alpha/R$, $\alpha$ a  constant), we achieve a rotating fluid without heat flux with all pressures equal to zero. That case corresponds to Kerr's solution.  
 We will generate rotating wormholes from statics wormholes via the NJA by imposing some geometric and physics conditions on $K$ and $\Psi$ functions; it will be the objective of the following section.
   \section{NJA for rotating wormholes} \label{sec:NJARW}
  \subsection{A class of static wormholes fixed by the Newman-Janis Ansatz}
To construct static wormholes from solutions of the Einstein field equations for an interior region of the space-time,  we first bring the metic (\ref{eq:GeneralStaticMetric}) to the form:
                           \begin{align}\label{eq:GSM_wormholes}
                                            \mathbf{ {\cal G}_0} &= - \e^{2\varphi(R)} \operatorname{d}{T} \otimes  \operatorname{d}{T} 
                                            + \frac{\operatorname{d}{R} \otimes  \operatorname{d}{R}}{1 -  {b(R)}/{R}}  
                                            +  R^2 ( \operatorname{d}{\Theta} \otimes  \operatorname{d}{\Theta}   
                                           + \sin^2 {\Theta} \operatorname{d}{\Phi} \otimes  \operatorname{d}{\Phi} ) \ .
                          \end{align} 
Here, we introduced the usual spherical coordinates $(R, \Theta, \Phi)$   and  performed the identification
                         \begin{align}\label{eq:metric_funtions_wormholes}
                                             G(R) = e^{ 2\varphi(R) }\ , \quad F(R) = 1 -  \frac{b(R)}{R}  \ , \quad  H(R) = R^2 .
                           \end{align}
The functions $\varphi$ and $b$ determine the gravitational redshift and the spatial shape of the wormhole \cite{morris1988wormholes}. Naturally, the match of wormhole to an exterior space-time must occur on a determined value $R_S$ of the radial coordinate $R$.  To guarantee that the Birkhoff theorem is satisfied in the wormhole exterior  is mandatory to impose the conditions:
          \begin{enumerate}
                   \item $ b(R) = b (R_S)  = \text{constant} \equiv b_0, R  >  R_S \ ,$
                   \item $\varphi(R) = \frac{1}{2} \ln(1 - {b_0}/{R}) , R  >  R_S  \ .$
              \end{enumerate}               
Additionally, to ensure that we get asymptotically flat solutions, we demand that:
               \begin{enumerate}
                   \item  $\lim_{R  \to \infty  } [{b(R)}/{R}]    \to 0 \ ,$
                   \item $\lim_{R \to  \infty } \varphi \to 0 \ .$
              \end{enumerate}                    
To embed, in the three-dimensional Euclidean space, a two-dimensional surface with the same geometry,  we set up $( \Theta = \pi/2, T = \text{constant} )$  in the metric (\ref{eq:GSM_wormholes})
\begin{align}
  \mathbf{ {\cal S}} & =   \frac{1}{1- {b(R)}/{R}}  \operatorname{d}{R} \otimes  \operatorname{d}{R}  +  R^2 \operatorname{d}{\Phi} \otimes  \operatorname{d}{\Phi}  \ .
\end{align} 
 Then, the surface of rotation
\begin{align}
{\cal \chi}(R, \Phi) =   (R \cos(\Phi) \ ,\ R \sin(\Phi) \ , Z(R) ) \ .
\end{align} 
  determines the spatial geometry of the wormhole spacetime  through the single function $Z(R)$, which satisfies the equation
   \begin{align}
     \frac{dZ(R)}{dR} =\pm\left( \frac{R}{b(R)}   -1 \right)^{-1/2} \ .
    \end{align} 
If we assume that metric  (\ref{eq:GSM_wormholes}) determines the existence of a static wormhole, their rotating counterpart generated by the NJA presents a restriction given by the  relation between the metric functions  provided by   $K^2 = ({{F}/{G}}) R^4$.  Hence, the unique solutions to the Einstein field equations describing static wormholes suitable to generate rotating wormholes by using NJA satisfies the relation:
              \begin{align}
                 e^{2\varphi(R)} = \frac{R^4}{K^2}{\left[ 1 - \frac{b(R)}{R}\right]}         
               \end{align} 
with $K$ determined by an arbitrary static solution of the Einstein equations, consistent with the five Classes in the former section. 

\subsection{Space-time for rotating wormholes}
The procedure employed here to construct rotating wormholes starts by bringing the stationary axially symmetric spacetime previously generated by the NJA  to the following form.
                                 \begin{align}\label{eq:Metric_RotWormholes}
                                             \mathbf{ {\cal G}} = & - e^{2\nu(R, \Theta)} \operatorname{d}{T} \otimes  \operatorname{d}{T} 
                                            + e^{2\mu_2 (R, \Theta)} \operatorname{d}{R} \otimes  \operatorname{d}{R} 
                                              +   e^{2\mu_3(R, \Theta)} \operatorname{d}{\Theta} \otimes  \operatorname{d}{\Theta} \nonumber\\ 
                                              & + e^{2\lambda(R, \Theta)} (\operatorname{d}{\Phi}  -  \omega \operatorname{d}{T}) \otimes     
                                             (\operatorname{d}{\Phi}  -  \omega \operatorname{d}{T})   \ ,
                         \end{align}
              where           
                     \begin{align*} 
                                         e^{2 \nu} &  = \frac{\Delta \Psi}{\rho^4}   \frac{[a^2 + K - a^2\sin^2\theta]^2}{(a^2 + K)^2 - a^2\sin^2\Theta\Delta} \ , \quad e^{2\mu_2} =\frac{\Psi}{\Delta} \ , \quad e^{2\mu_3} = \Psi \ , \\
                                         e^{2\lambda} & = \sin^2\Theta\frac{\Psi}{\rho^4} [( a^2 + K)^2 - a^2\sin^2\Theta \Delta] \ , \quad  \omega = \frac{a(a^2 + K - \Delta)}{ (a^2 + K)^2 - a^2\sin^2\theta \Delta} \ , \\
                                         K^2 &  = \frac{F}{G}H^2 \ , \quad \rho^2 = K + a^2\cos^2\Theta \ , \quad  \Delta = FH + a^2 \ 
                                                \end{align*}   
where, as we know, each couple  ($K \ , \Psi)$ sets a new rotating solution of the Einstein field equations provided a static solution is known.  Next,  we suppose that there exist a function $B= B(R, \Theta) $ such that \cite{teo1998rotating, azreg2016wormholeA, azreg2016wormholeB}:
      \begin{align*}
                  e^{2\mu_2} = \mathbf{ {\cal G}}_{RR} = \frac{1}{1 - B/R} \ , \qquad
                        \lim_{R \to \infty } \frac{B}{R}  = 0 \ , \qquad
                  \lim_{R \to \infty }  \mathbf{ {\cal G}}_{TT}  = 1 \  . 
                  \end{align*} 
Additionally, to avoid conic singularities  we  assume that,  on the axis of rotation $ (\Theta =0  \ \text{or} \  \Theta = \pi)$,
 the following relation it is always possible:
               \begin{align*}
	      { \mathbf{\cal G}  }_{\Theta\Theta}   =  \frac{{\mathbf{ \cal G}}_{\Phi \Phi}}{\sin^2\Theta}   
	                     \end{align*} 
and, that near the throat : 
   \begin{align*} R\, \partial_{R}B  < B  .
       \end{align*}   
Furthermore, to ensure the applicability of the NJA, the choices of the metric functions corresponding to the static space-time and the corresponding to the rotating space-time 
are restricted by the following constrain.
    \begin{align}
    F =  1  -  \frac{b(R)}{R}\ ,\qquad 
      e^{2\varphi(R)} = \frac{R^4}{K^2}{\left[ 1 - \frac{b(R)}{R}\right]} \ ,
              \end{align}
where, as is known,  exists a unique correspondence between each choice of $K$ and $\Psi$ determined by the classes of rotating solutions discussed above.  Then, finally, the weakest condition to be satisfied by the rotating wormhole is given by the following expression   
      \begin{align}
      \frac{B(R, \Theta)}{R} = 1 -  \frac{R^2}{\Psi}\left( 1 -  \frac{b(R)}{R}  + \frac{a^2}{R}  \right) ,
            \end{align}
provided the $b(R)$  function defines a static wormhole and  $B(R, \Theta)$ satisfies all the conditions previously discussed.
 	   
 \subsection{Example of Rotating wormholes from the NJA}
As a straightforward example, let us consider the static solution
      \begin{align}
      b(R) =R_S  - \frac{q^2}{R}\ , 
        \end{align}
      where $R_S$ and $q$ are real constants.
      This solution determine the Reissner-Nordstr\"om wormhole,  the spatial geometry of the wormhole spacetime  through the single function $Z(R)$ satisfying
     \begin{align}
     dZ(R) =\pm\left( \frac{R}{R_S  - {q^2}/{R}}   -1 \right)^{-1/2}  d R \ .
     \end{align}
  and energy-momentum tensor given by   \cite{morris1988wormholes}
         \begin{align}
     \sigma(R) = p(R) = \tau(R) = \frac{q^2}{8\pi R^4} .
         \end{align}
Accordingly, if we choose, for example $K$ and $\Psi$  as given by the Class 2 of rotating solutions obtained by the  NJA, then we achieve a Kerr-Newman wormhole:
      \begin{align*}
      \frac{B(r, \theta)}{R} = 1 -  \frac{R^2}{R^2 + a^2\cos^2\Theta}\left( 1 -  \frac{q^2}{R}  + \frac{a^2}{R}  \right) .
            \end{align*}
 The wormhole rotates with angular velocity given by
    \begin{align*}
    \omega= \frac{a R_S R}{(a^2 + R^2)^2 - a^2\sin^2\theta(a^2 + R^2 + q^2 - R_S R)}\ ,
    \end{align*}
 and their corresponding energy-momentum tensor is given by   
    Dynamical quantities:
    \begin{align*}
    \sigma = \tau= p=  \frac{q^2}{8\pi (a^2\cos^2\theta + R^2)^2} \ .
    \end{align*}
In other words, the rotating counterpart of the Reissner-Nordstr\"om wormhole obtained by the NJA corresponds to the Kerr-Newman wormhole. 
 \section{Conclude remarks}
  This paper presented the ansatz formulated by Newman-Janis to generate solutions to the Einstein field equations for rotating space-times. We performed the algorithm's structure based on modern differential geometry to keep open the discussion about their nature. Actuality, We discuss the algorithm steps setting successive maps on the tangent and cotangent spaces, and compile some independent classes of solutions of the Einstein field equations for rotating space-times, all obtained from prescribed static and spherically space-times.
  
We have elaborated on the possibility of constructing static wormholes and their rotating counterpart by choosing any of the classes of solutions generated by the Newnan-Janis Ansatz. These classes restrict the nature of both static and rotating wormholes.
To illustrate the discussed ideas,  we generated the Kerr-Newman wormholes using the Reissner-Nordstr\"om wormhole as a seed solution.  
It will be fructiferous to produce many examples of rotating wormholes operating with the ideas presented in this work. The problem of match wormholes to their exterior space-time remains open. One emergent here is studying the wormholes' asymptotically flat characteristics and using a $C^3$  junction technique \cite{gutierrez2019c3, gutierrez2021darmois}. We will explore that in subsequent work.
  \section*{Acknowledgments} 
   A.C.G-P is grateful for the invitation to participate in XXII International Meeting
Physical Interpretations of Relativity Theory - 2021 (Moscow, Russia) .
  This work was partially supported by Programa Capital Semilla para Investigaci\'on, Proyecto 2490, VIE-UIS .


\section*{References}
\begin{thebibliography}{10}
\expandafter\ifx\csname url\endcsname\relax
  \def\url#1{{\tt #1}}\fi
\expandafter\ifx\csname urlprefix\endcsname\relax\def\urlprefix{URL }\fi
\providecommand{\eprint}[2][]{\url{#2}}

\bibitem{kerr1963gravitational}
Kerr R~P 1963 {\em Physical review letters\/} {\bf 11} 237

\bibitem{newman1965note}
Newman E~T and Janis A 1965 {\em Journal of Mathematical Physics\/} {\bf 6}
  915--917

\bibitem{drake2000uniqueness}
Drake S and Szekeres P 2000 {\em General relativity and Gravitation\/} {\bf 32}
  445--457

\bibitem{azreg2014static}
Azreg-A{\"\i}nou M 2014 {\em The European Physical Journal C\/} {\bf 74} 2865

\bibitem{gutierrez2016newman}
Guti{\'e}rrez-Pi{\~n}eres A~C and Quevedo H 2016 {\em General Relativity and
  Gravitation\/} {\bf 48} 146

\bibitem{ellis1973ether}
Ellis H~G 1973 {\em Journal of Mathematical Physics\/} {\bf 14} 104--118

\bibitem{morris1988wormholes}
Morris M~S and Thorne K~S 1988 {\em American Journal of Physics\/} {\bf 56}
  395--412

\bibitem{lemos2003morris}
Lemos J~P, Lobo F~S and de~Oliveira S~Q 2003 {\em Physical Review D\/} {\bf 68}
  064004

\bibitem{teo1998rotating}
Teo E 1998 {\em Physical Review D\/} {\bf 58} 024014

\bibitem{azreg2016wormholeA}
Azreg-Ainou M 2016 {\em The European Physical Journal C\/} {\bf 76} 3

\bibitem{azreg2016wormholeB}
Azreg-Ainou M 2016 {\em The European Physical Journal C\/} {\bf 76} 7

\bibitem{de1992relativity}
De~Felice F and Clarke C~J~S 1992 {\em Relativity on curved manifolds\/}
  (Cambridge University Press)

\bibitem{gutierrez2019c3}
Guti{\'e}rrez-Pi{\~n}eres A~C and Quevedo H 2019 {\em Classical and Quantum
  Gravity\/} {\bf 36} 135003

\bibitem{gutierrez2021darmois}
Guti{\'e}rrez-Pi{\~n}eres A~C and Quevedo H 2021 {\em arXiv preprint
  arXiv:2106.08679\/}

\end{thebibliography}
\end{document}